\makeatletter

\newcommand{\Rmnum}[1]{\expandafter\@slowromancap\romannumeral #1@}
\makeatother

\documentclass[conference]{IEEEtran}
\IEEEoverridecommandlockouts
\usepackage{cite}
\usepackage{amsmath,amssymb,amsfonts}
\usepackage{algorithmic}
\usepackage{graphicx}
\usepackage{textcomp}
\usepackage{xcolor}
\usepackage{float}
\usepackage{lscape}
\usepackage{booktabs}
\usepackage{algorithm}
\usepackage{fancyhdr} 

\def\BibTeX{{\rm B\kern-.05em{\sc i\kern-.025em b}\kern-.08em
    T\kern-.1667em\lower.7ex\hbox{E}\kern-.125emX}}

\newcommand{\noteblue}[1]{\textcolor{blue}{[{\bf #1}]}}

\begin{document}

\title{Deep Reinforcement Learning-based Radio Resource Allocation and Beam Management under Location Uncertainty in 5G mmWave Networks  \\
}

\author{\IEEEauthorblockN{Yujie Yao, Hao Zhou, and Melike Erol-Kantarci, \IEEEmembership{Senior Member, IEEE}}
\IEEEauthorblockA{\textit{School of Electrical Engineering and Computer Science} \\
\textit{University of Ottawa}\\
Emails:\{yyao016,hzhou098, melike.erolkantarci\}@uottawa.ca}}

\maketitle

\thispagestyle{fancy} %
      \lhead{} 
      \chead{Accepted by 2022 IEEE Symposium on Computers and Communications, \copyright2022 IEEE } 
      \rhead{} 
      \lfoot{} 
      \cfoot{\thepage} 
      \rfoot{} 
      \renewcommand{\headrulewidth}{0pt} 
      \renewcommand{\footrulewidth}{0pt} 
\pagestyle{fancy}

\begin{abstract}
Millimeter Wave (mmWave) is an important part of 5G new radio (NR), in which highly directional beams are adapted to compensate for the substantial propagation loss based on UE locations. However, the location information may have some errors such as GPS errors. In any case, some uncertainty, and localization error is unavoidable in most settings. Applying these distorted locations for clustering will increase the error of beam management. Meanwhile, the traffic demand may change dynamically in the wireless environment. Therefore, a scheme that can handle both the uncertainty of localization and dynamic radio resource allocation is needed. In this paper, we propose a UK-means-based clustering and deep reinforcement learning-based resource allocation algorithm (UK-DRL) for radio resource allocation and beam management in 5G mmWave networks. We first apply UK-means as the clustering algorithm to mitigate the localization uncertainty, then deep reinforcement learning (DRL) is adopted to dynamically allocate radio resources. Finally, we compare the UK-DRL with K-means-based clustering and DRL-based resource allocation algorithm (K-DRL), the simulations show that our proposed UK-DRL-based method achieves 150\% higher throughput and 61.5\% lower delay compared with K-DRL when traffic load is 4Mbps.
\end{abstract}

\begin{IEEEkeywords}
Beam management, UK-means, Localization Uncertainty, Radio Resource allocation, Deep reinforcement Learning
\end{IEEEkeywords}

\section{Introduction}
Fifth-generation (5G) network is characterized by multi-scenario applications, a considerable number of devices and huge data volume. Millimeter Wave (mmWave) communication is developed to overcome a number of obstacles of mobile networks, especially the high data rate challenge and low latency requirement \cite{b1}. It provides users with rich bandwidth which allows for a higher transmission rate. However, one vital shortcoming of mmWave is the substantial propagation loss, which greatly limits its coverage range. 

To this end, highly directional beams which can provide large antenna array gains with less inter-beam interference are adopted to solve this problem \cite{b2}. Beam management is performed to align the beam pairs between user equipment (UE) and base station (BS). A typical way of beam management is to have all users divided into numerous clusters, and each cluster is served by a separate beam \cite{B2}. 
In 5G, the network traffic will experience rapid change, which prompts BS to dynamically adjust beams to cover active UEs. Consequently, the UE localization becomes a key pre-requisite for clustering. There are various technologies proposed for outdoor localization with the assistance of BS to replace traditional global positioning system, such as ToA positioning method \cite{b3} and AoA-based positioning techniques \cite{b4}.
Although each algorithm has its own advantages, they all bring localization errors, which will decrease the performance of the network. 
As such, a radio resource allocation method that can handle both the localization and traffic demand uncertainty of users need to be developed. 

On the other hand, the wireless network nowadays is faced with heterogeneous service requirements, and machine learning becomes an ideal solution for intelligent and flexible allocation of network resources \cite{B4_3}. Specifically, reinforcement learning (RL) is designed to exploit the current environment without any prior knowledge, which avoids the complexity of building a dedicated optimization model\cite{B4_4}. Moreover, combined with a deep neural network, deep reinforcement learning (DRL) overcomes the limitations of tabular-based RL,  which has been generally applied for wireless network optimizations \cite{B_zh}.

In this paper, we consider a 5G NR network using mmWave which serves multiple single-beam UEs, and we proposed a UK-means-based clustering and DRL-based resource allocation algorithm (UK-DRL). Firstly, we apply the UK-means algorithm to tackle the localization uncertainty\cite{b5}. More specifically, instead of the exact position, UK-means uses the probability distribution function (PDF) of objects for clustering, which is designed to handle the clustering with data uncertainty. Then, we employ a DRL-based algorithm for radio resource allocation, which jointly considers the Quality-of-Service (QoS) requirements of different types of users \cite{b6}. We compare our proposed method with K-means-based clustering and DRL-based resource allocation algorithm (K-DRL) that clusters UEs with localization uncertainty. Our results show that our proposed method has 150\% higher throughput and 61.5\% lower delay under high load. Meanwhile, we feed K-DRL with exact data as an ideal optimal baseline, and simulations show that our UK-DRL presents a comparable performance with the ideal results where exact location information is available.

The rest of this paper is organized as follows. Section \ref{s2} discusses the related work, and Section \ref{s3} introduces the system model. 
Section \ref{s4} describes the clustering methods and DRL algorithm. Section \ref{s5} presents simulation settings and results. Lastly, Section \ref{s6} concludes the paper.

\section{Related Work}
\label{s2}

Localization and clustering are generally considered as an important part of resource allocation, since the network performance is directly related to UE locations.  In \cite{b6}, Density-Based Spatial Clustering of Applications with Noise (DBSCAN) is adopted to cluster users, then a DRL algorithm is introduced to allocate radio resource. Considering the Non-orthogonal multiple access (NOMA) system, a hierarchical clustering scheme is proposed in \cite{b7} to obtain the optimal number of clusters, which enables the BS to form the optimal steering direction of mmWave beams and achieves higher throughput. Two schemes of location-aided power allocation optimization are proposed in \cite{b8}. The authors firstly determined the optimal power allocation to each cluster and then optimized the intra-beam power allocation. Authors of \cite{b9} developed a power allocation policy to maximize the sum rate in a millimeter-wave non-orthogonal NOMA system, and a K-means-based clustering algorithm is applied to reduce the computational complexity.

Most aforementioned studies assume the BS can observe the exact location of UEs, then they implement the beamforming and resource allocation accordingly. However, this is an unrealistic assumption in practice, and the localization uncertainty is unavoidable due to obstacles, interference and so on. Different than previous studies, we include the localization uncertainty of UEs in this paper. The UK-means  algorithm is implemented to handle the uncertainty of UE locations. The algorithm uses distribution functions for clustering. Compared with traditional clustering algorithms such as K-means, UK-means can obtain clustering results that are more consistent with the real locations.

\section{System Model}
\label{s3}
\subsection{Network Model}
\begin{figure}[t!]
    \centering
    \includegraphics[width=0.4\textwidth]{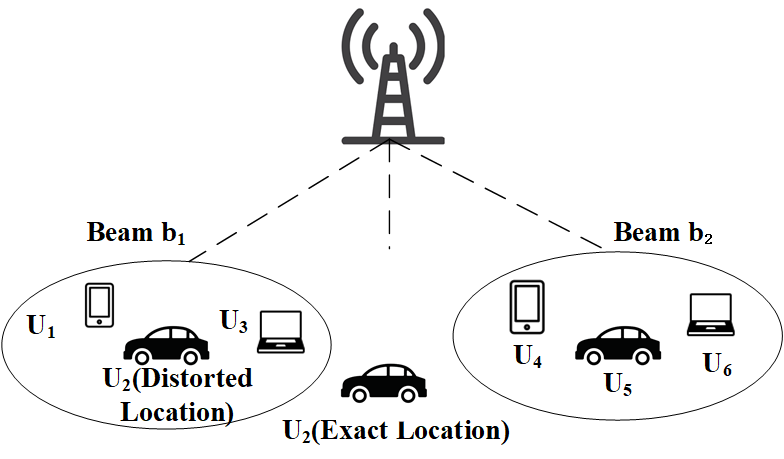}
    \caption{System model of mmWave networks with highly directional beams.} 
    \label{fig1}
\end{figure}

As shown in Fig. \ref{fig1}, here we assume there is a gNB that serves several users denoted by \textbf{\emph{U}}, and each user is represented by $U_i$. The gNB will observe the location information of users, and perform clustering to group the users. Based on the observed location, each cluster is served by a separate beam denoted by $b_i\in{\boldsymbol{B}}$.  

However, the UE $U_2$ suffers from a localization error, which may be caused by band channel conditions or UE mobility. The exact location of $U_2$ is between two formed clusters, which means $U_2$ is not served by any of the two formed beams. As a result, the transmission of $U_2$ may experience a long delay, which proves that the localization uncertainty affects the network performance. In this work, we address this problem by applying a UK-means clustering algorithm, which aims to mitigate the effect of localization uncertainty.

Moreover, in the resource allocation phase, we assume each beam serves its users via Orthogonal Frequency Division Multiple Access (OFDMA), and intra-beam interference can be avoided. The available bandwidth $w_i$ for beam $b_i$ is split into Resource Blocks (RBs), each of which has 12 subcarriers. A Resource Block Group (RBG) is formed by combining contiguous RBs together, which is considered as the smallest radio resources that can be allocated \cite{b6}.

\subsection{Channel Model}
Here we assume the gNB is equipped with $N_t$ antennas. Compared with the gain of Line-of-Sight (LoS) path, the gain of non-Line-of-Sight (nLoS) paths can be neglected. As a result, the mmWave channel can be modeled using a single LoS path model \cite{b9}. According to \cite{b12}
, the mmWave channel between the gNB and the $u^{th}$ UE can be represented by
\begin{equation}
    \boldsymbol{H}_u=\sqrt{\frac{N_t}{M_u}}\sum_{l=1}^{M_u}\alpha_{u,l}\boldsymbol{a}_u(\theta_{u,l})\boldsymbol{a}_{gNB}(\varphi_{u,l})
\end{equation}
where $M_u$ denotes the number of paths between gNB and UE, $\alpha_{u,l}$ is the complex gain of $l^{th}$ path. $\theta_{u,l}$ denotes the angle of arrival (AOA) and $\varphi_{u,l}$ denotes the angle of departure (AOD). Moreover, $\boldsymbol{a}_u(\theta_{u,l})$ and $\boldsymbol{a}_{gNB}(\varphi_{u,l})$ are the antenna array response vectors of the gNB and $u^{th}$ UE respectively.

If a Uniform Linear Array (ULA) is considered, 
$\boldsymbol{a}_{gNB}(\varphi_{u,l})$ can be represented by
\begin{equation}
    \boldsymbol{a}_{gNB}(\varphi)=\frac{1}{\sqrt{N_t}}[1,e^{j2\pi\frac{d}{\lambda}\sin({\varphi})},...,e^{j(N_t-1)2\pi\frac{d}{\lambda}\sin({\varphi})}]^T
\end{equation}
where $d$ denotes the distance between gNB's antennas and $\lambda$ denotes the wavelength.

\section{UK-means-based Deep Reinforcement Learning for Radio Resource Allocation and Beamforming}
\label{s4}
To handle localization uncertainty in clustering, we adopted the UK-means for clustering \cite{b5}. The clustering results will be transformed into a set of beams with UEs they serve. Then the DRL is adopted in each beam to allocate physical resource blocks.

\subsection{ UK-means-based Clustering and Beamforming}
The classical K-means algorithm consists of 4 steps:\\
\indent Step 1: Assign initial centers of clusters.\\
\indent Step 2: Compute the distance between the data and centers, assign them to the cluster with minimum distance.\\
\indent Step 3: Recalculate cluster centers.\\
\indent Step 4: Repeat 2-3 until the algorithm converges.

The UK-means algorithm follows the same steps as K-means. But the differences are that it applies different formulas to calculate distance and update cluster centers with the assistance of PDF. UK-means enables us to get the most similar clustering results with distorted position information compared to the results with exact position information. 

Let us assume we need to form $n$ clusters, and the $j^{th}$ cluster is denoted by $C_j$. The center of the cluster is denoted by $\boldsymbol{c_j}$. For each user position $\boldsymbol{x_i}$, the uncertainty is represented by a known PDF $f(\boldsymbol{x_i})$.

In K-means, the Euclidean distance between data point $\boldsymbol{x_i}$ and cluster center $\boldsymbol{c_j}$ is calculated by:
\begin{equation}
    ||\boldsymbol{x_i}-\boldsymbol{c_j}||=\sqrt{\sum_{i=1}^V |x_i-c_j|^2},
    \label{eq3}
\end{equation}
where $V$ represents the dimension of data. In this paper, we consider two-dimensional space.

The center point update formula is as follows:
\begin{equation}
    \boldsymbol{c_j}=\frac{1}{C_j}\sum_{i\in{C_j}}\boldsymbol{x_i}
    \label{eq4}
\end{equation}
where $|C_j|$ represents the number of users that are assigned to cluster $|C_j|$.

The two formulas applied in UK-means are as follows \cite{b5}:
\begin{equation}
    E(||\boldsymbol{x_i}-\boldsymbol{c_j}||^2)=\int{ ||\boldsymbol{x_i}-\boldsymbol{c_j}||^2f(\boldsymbol{x_i}) d{\boldsymbol{x_i}}}
    \label{eq5}
\end{equation}

\begin{equation}
    \boldsymbol{c_j}=\frac{1}{|C_j|}\sum_{i\in{C_j}}\int{\boldsymbol{x_i}f(\boldsymbol{x_i})d{\boldsymbol{x_i}}}
    \label{eq6}
\end{equation}

In equations (\ref{eq5}) and (\ref{eq6}), the distance is an expected distance between the cluster center and uncertain data. It is computed from the double integral over the uncertainty region. The center of one cluster is computed as the mean of the expected values over the PDFs of the uncertain data in the same cluster.

In this paper, we assume the uncertainty region is specified as a circle with center $\boldsymbol{x_i}$ and radius $R$. Node positions are distributed uniformly within the circle. To make the calculation easier, polar coordinate system is applied. The PDF for a uniformly distributed circle is:
\begin{equation}
    f(r,\theta)=\frac{r}{\pi{R^2}},
\end{equation}
where $r\in{[0,R]}$ and $\pi\in{[0,2\pi]}$.

In this way, equation (\ref{eq5}) can be rewritten as:
\begin{equation}
    E(||\boldsymbol{x_i}-\boldsymbol{c_j}||^2)=\int_{0}^{R}{\int_{0}^{2\pi}{ ||\boldsymbol{x_i}-\boldsymbol{c_j}||^2f(r,\theta) rd{r}d{\theta}}}
\end{equation}

For a data point $\boldsymbol{x_i}$, its Cartesian coordinate is $(x_i,y_i)$. Assume the coordinate of center $\boldsymbol{c_j}$ is $(c_x^j,c_y^j)$, then:
\begin{equation}
    \begin{aligned}
        ||\boldsymbol{x_i}-\boldsymbol{c_j}||^2&=(x_i+r\cos\theta-c_x^j)^2+(y_i+r\sin\theta-c_x^j)^2\\
        &=2r\sin\theta(x_i-c_x^j)+2r\cos\theta(y_i-c_y^j)\\
        &+(x_i-c_x^j)^2+(y_i-c_y^j)^2+r^2,
    \end{aligned}
    \label{eq9}
\end{equation}
where $(x_i+r\cos\theta,y_i+r\sin\theta)$ denotes the distorted location. The error is represented in the form of cylindrical coordinate system.

\subsection{Deep Reinforcement Learning-based Radio Resource Allocation}

\begin{figure}
    \centering
    \includegraphics[width=0.4\textwidth]{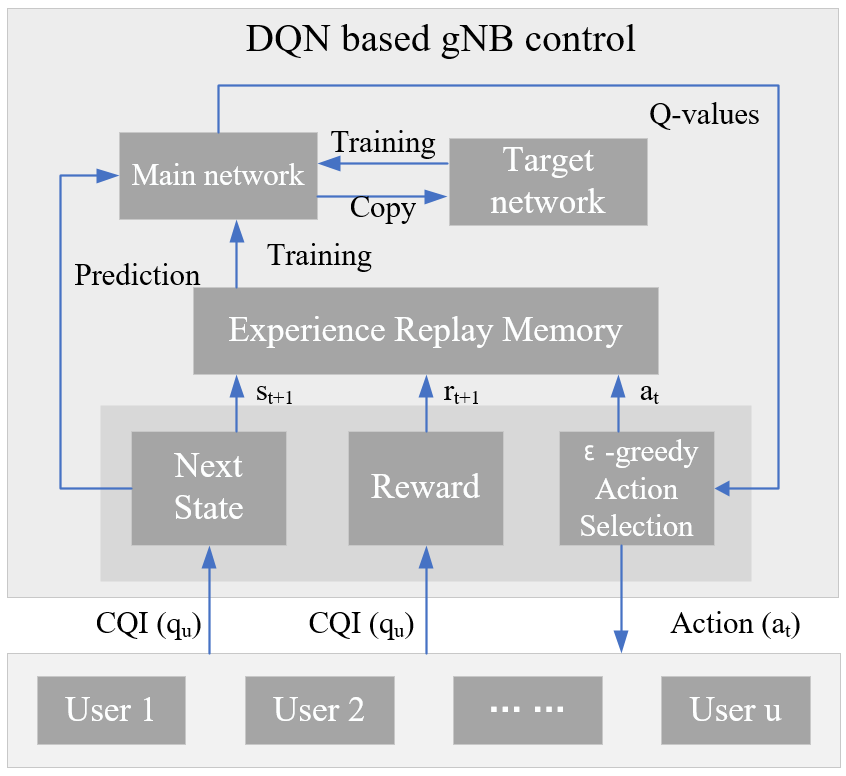}
    \caption{Diagram of LSTM-based Deep Q-learning.
    \label{fig2}
}
\end{figure}
After the UK-means-based clustering, we assume each mmWave beam will cover a cluster of UEs within a specific angle. Then we applied the DRL algorithm  in each beam to allocate radio resources to intra-beam UEs, aiming to maintain high QoS requirements for both ultra-reliable low latency communications (URLLC) and enhanced mobile broadband (eMBB) users. The Markov decision process of DRL is defined as follows:

1) Actions: In $b^{th}$ beam, the action is denoted by allocating $i^{th}$ RBG to $j^{th}$ user, which is defined by
\begin{equation}
    a_i^b = {u_j^b}
\end{equation}

2) States: States are specified as the channel quality indicator (CQI) feedback provided by the user. As a result, the state of $i^{th}$ RBG in the $b^{th}$ beam can be expressed as
\begin{equation}
    s_i^b = {q_i^b}
\end{equation}
where $q_i^b$ is the CQI of $i^{th}$ RBG in the $b^{th}$ beam.

3) Reward: The reward function needs to consider the QoS requirements for both URLLC and eMBB users. URLLC users require high reliability and low latency, while eMBB users desire a high data rate. The reward function of $i^{th}$ RBG in the $b^{th}$ beam can be written as:
\begin{equation}
    r_{b,i}=
    \begin{cases}
    sigm({\frac{S_i}{S^{QoS}}}),& ~eMBB ~user\\
    sigm(\frac{S_i}{S^{QoS}} \frac{T^{QoS}}{T_{u}^{q}} ),& ~URLLC~ user 
    \end{cases}
    \label{eq12}
\end{equation}
where $S_i$ is the signal-to-noise ratio (SINR) of the link allocation of $i^{th}$ RBG to a user in the $b^{th}$ beam, $S^{QoS}$ is the SINR requirement of eMBB users, $T^{QoS}$ is the latency requirement of URLLC users, and $T_{u}^{q}$ is the queuing delay.
$sigm$ means the sigmoid function, which will keep the reward between the interval [0,1]:
\begin{equation}
    sigm(x)=\frac{1}{1+e^{-x}}
\end{equation} 

In equation (\ref{eq12}), the reward of eMBB users depends on the SINR ratio, which means a higher throughput will lead to a higher reward. On the other hand, the URLLC reward consists of two terms, in which the $\frac{T^{QoS}}{T_{u}^{q}}$ indicates that a lower delay will bring a higher reward, and $\frac{S_i}{S^{QoS}}$ indicates the requirement for higher reliability.

Fig. \ref{fig2} shows the diagram of DRL-based intra-beam radio resource allocation. In the $t^{th}$ time interval, the current state and action are represented by $s_t$ and $a_t$, respectively. The agent will first select an action based on $\epsilon$-greedy policy, and the actions are implemented in the wireless environment. Then the next state $s_{t+1}$ and reward $r_{t+1}$ are calculated with the assist of CQI and SINR from the users. The $s_t$, $a_t$, $s_{t+1}$ and $r_{t+1}$ forms an experience tuple, which is saved in the experience pool. The experience will be further used to train the neural network which will predict the Q-values. Lastly, the agent will determine the next action $a_{t+1}$ based on predicted Q-values.

In this work, we apply long short-term memory (LSTM) network for Q-values prediction. As a special recurrent neural network, LSTM can better capture the long-term data dependencies than traditional neural networks, which makes it a favourable choice to handle the complicated wireless environment. A pack of experience tuples $<s_t, a_t, s_{t+1}, r_{t+1}>$ are sampled from the experience pool after several iterations, which will be used to train the main network by gradient descent algorithm: 
\begin{equation}
L(w)=Er(r_{t}+\gamma \max\limits_{a} Q(s_{t+1},a,\theta')-Q(s_{t},a_{t},\theta))
\label{eq14}
\end{equation}

where $\theta$ and $\theta'$ represent the weight of main and target LSTM networks, respectively. $Er()$ indicates the error function. $r^{t}+\gamma \max\limits_{a} Q(s^{t+1},a,\theta')$ shows that target network will predict the target Q-values, and $Q(s^{t},a^{t},\theta)$ shows main network will produce current Q-values.
Then the network weight will be copied to the target network after several training iterations. Such late update of the target network will provide a stable reference for the main network.

The steps of the proposed UK-DRL algorithm are presented in \textbf{Algorithm 1}.

\begin{figure}
    \centering
    \includegraphics[width=0.4\textwidth]{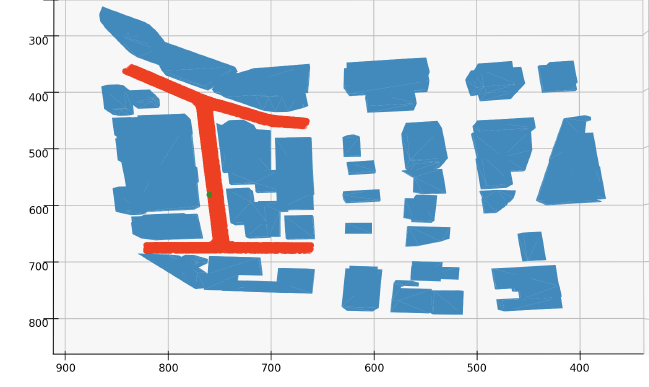}
    \caption{Map of Rosslyn area in the applied dataset \cite{B12_2}.
    \label{map}
}
\end{figure}

\begin{algorithm}
	\renewcommand{\algorithmicrequire}{\textbf{Input:}}
	\renewcommand{\algorithmicensure}{\textbf{Output:}}
	\caption{UK-DRL}
	\label{alg1}
	\begin{algorithmic}[1]
		\STATE \textbf{Initialization:} Wireless and learning parameters.
		\FOR{TTI t = 1 to T}
		\STATE Updating clustering queue and get location of users
		\STATE Performing UK-means using equation (\ref{eq6}) and (\ref{eq9}). Generating beams based on clustering results.
		\FOR{Each beam $b_i\in{\boldsymbol{B}}$}
		\STATE Selecting an action $a_t$ in each beam based on $\epsilon$-greedy algorithm.
		\FOR{User $u_j\in{b_i}$}
		\STATE Perform the action assigned by gNB.
		\STATE Send uplink report (i.e. SINR) to gNB.
		\ENDFOR
		\STATE  Compute the reward $r_{t+1}$ in equation (\ref{eq12}).
		\STATE  Store the experience tuple in the experience replay memory.
		\STATE  Sample randomly from experience replay memory every $C$ TTIs and apply equation (\ref{eq14}).
		\STATE  Predict the Q-values for next action selection.
		\ENDFOR
		\ENDFOR
	\end{algorithmic}  
\end{algorithm}

\section{Simulation Settings and Results}
\label{s5}
\subsection{Simulation Settings}

In the simulations, we consider an environment shown by Fig. \ref{map}, which is given by a realistic dataset Raymobtime s009 \cite{B12_1}. The data is generated according to Rosslyn, an urban district in Arlington, Virginia. The blue regions represent buildings. The red data points reflect vehicle distribution on the streets. The dataset concludes episodes and scenes, where one episode is composed of a series of scenes. Ray-tracing data, lidar and camera images are included. The position information it provides is used as exact locations in our simulation.

\begin{table}
\centering
\caption{Simulation Parameters}
\begin{tabular}{|p{5cm}|p{1.5cm}|}
     \hline
     \textbf{Network Model}&  \\
     \hline
     Number of URLLC per cluster&1\\
     Number of eMBB per cluster&1\\
     Number of clusters&3\\
     Radius of cell&160m\\
     Beam angle&20$^\circ$\\
     $N_t$&1024\\
     \hline
     \textbf{Q-Learning}&  \\
     \hline
     Learning rate ($\alpha$)&0.5\\
     Discount factor ($\gamma$)&0.9\\
     Exploration probability ($\epsilon$)&0.1\\
     \hline
     \textbf{LSTM}&\\
     \hline
     Size of input layer&1\\
     Number of hidden units&20\\
     Size of output layer&24\\
     Size of mini-batch&20\\
     Size of replay memory&60\\
     Training Interval&60\\
     Copy Interval&120\\
     \hline
     \textbf{Simulation Parameters}&   \\
     \hline
     Simulation time&0.2s\\
     Number of TTIs in Every Run&1400\\
     Number of Runs&5\\
     Confidence Interval&95\%\\
     \hline
\end{tabular}
\label{table1}
\end{table}

Based on the dataset, we include one gNB and 6 UEs in our simulation. The radius of the gNB is 160 meters. The traffic of users follows Poisson distribution with $\lambda$ inter-arrival time and a packet size of 32 bytes. During our simulation, we assume three beams with 20$^\circ$ angle are formed. The QoS requirement of latency $T^{QoS}$ is equal to 1 ms, and the QoS requirement for SINR $S^{QoS}$ is set to 15 dB \cite{b13}. Other settings of DRL are shown in Table \ref{table1}. The total simulation time is 1400 TTIs. The simulation is repeated 5 times, and the confidence interval is 95\%.

We have three defined scenarios: 
\begin{itemize}
    \item Scenario 1: In this scenario, we adapt the predicted position given by \cite{B12_2} as locations with error. \cite{B12_2} proposed a cascaded image classification and deep learning technique to localize users using images and ray-tracing data. The root mean squared error is 8 m. We applied K-means to cluster UEs with location error, and DRL for resource allocation. The scenario is named by K-DRL+Location with Error.
    \item Scenario 2: Compared with scenario 1, the only difference in scenario 2 is that we use UK-means to group the UEs with localization error. The scenario is named by UK-DRL+Location with Error.
    \item Scenario 3: In scenario 3, we use the exact location provided by Raymobtime s009 dataset for clustering, which is considered as an optimal baseline. This scenario is named by K-DRL+Exact Location.
\end{itemize}

\begin{figure}
    \centering
    \includegraphics[width=8cm,height=5.2cm]{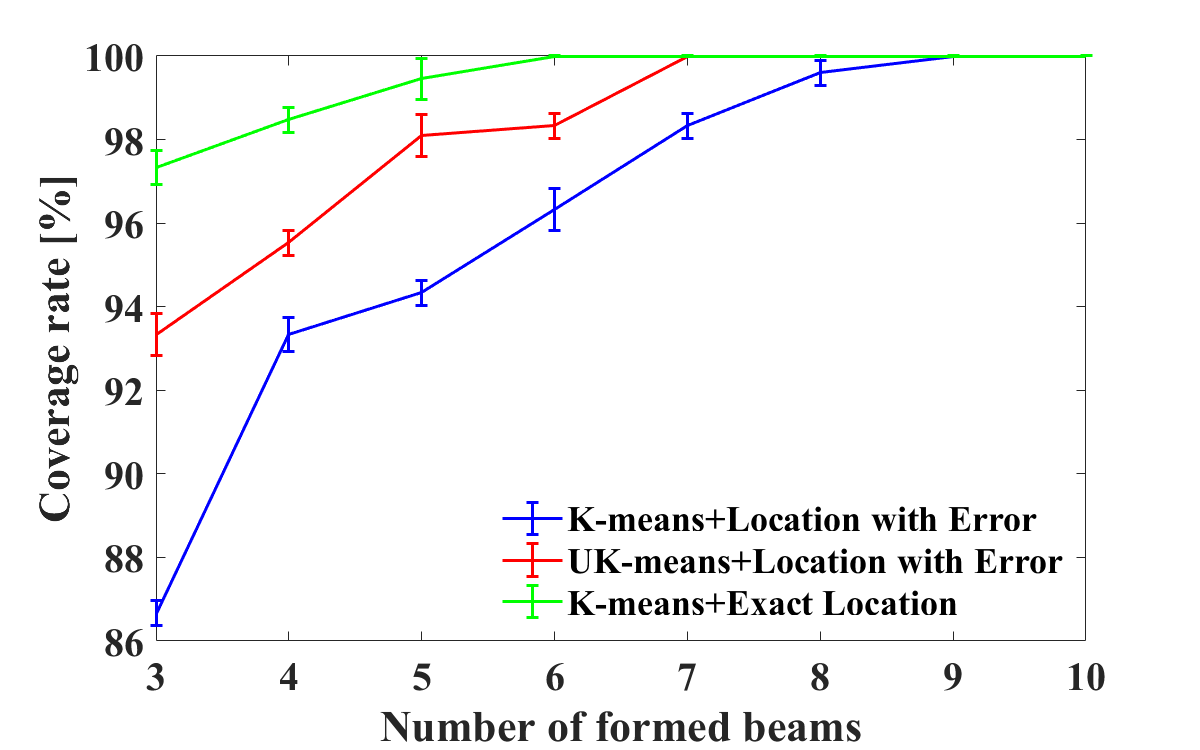}
    \caption{Coverage rate comparison under varying number of beams.
    \label{fig3}
}
\end{figure}
\begin{figure}
    \centering
    \includegraphics[width=8cm,height=5.2cm]{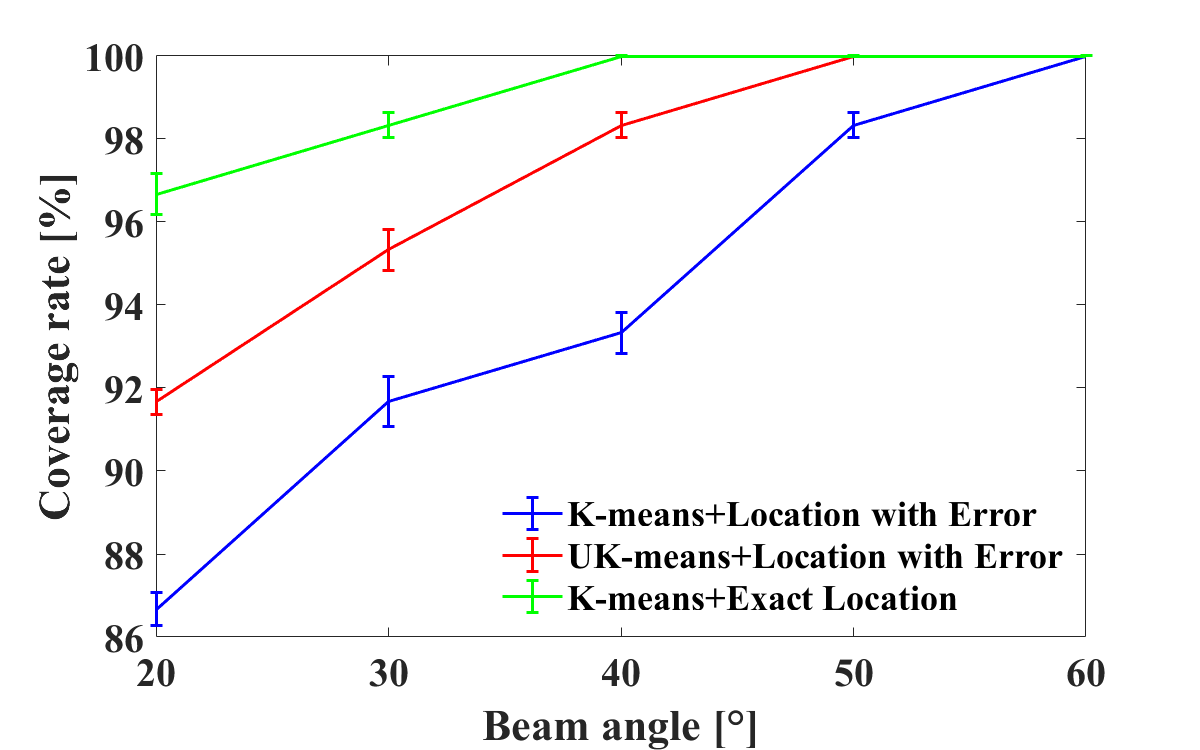}
    \caption{Coverage rate comparison under varying beam angle.
    \label{fig4}
}
\end{figure}

\subsection{Simulation Results}
\subsubsection{Results for Coverage Rate}
In our experiment, we firstly compare the number of beams to achieve full coverage and the average coverage rate under different beam angles when the number of beams is fixed to 4. 

Fig. \ref{fig3} shows the coverage rate under a different number of beams when beam angle is 30$^\circ$. In order to achieve full coverage, the third scenario needs 6 beams due to its exact locations. When distorted locations are applied, the UK-means approach requires 7 beams. However, the K-means-based method needs 9 beams to fully cover all the UEs. On the other hand, we investigate the coverage rate against different beam angles in Fig. \ref{fig4}. The beam number is fixed to 4 beams. Scenario 3 still has the best coverage rate because of the exact location. When distorted locations are applied, the localization error will decrease the coverage rate. Compared with K-means, UK-means has an improvement of 7\% when beam angle is 20$^\circ$. Fig. \ref{fig3} and \ref{fig4} demonstrate that we can achieve a better coverage rate with the same number of beams by using UK-means method when localization error exists.

\subsubsection{Results for Resource Allocation}
In this section, we display the simulation results of the three scenarios under different traffic loads. The network performance is evaluated by the total data rate and average delay.

\begin{figure}
    \centering
    \includegraphics[width=8cm,height=5.2cm]{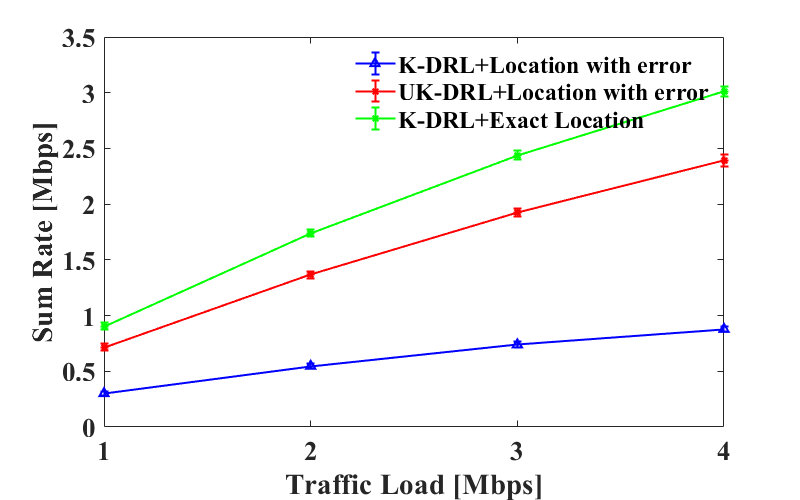}
    \caption{Sum rate comparison under varying traffic loads.
    \label{fig5}
}
\end{figure}

\begin{figure}
    \centering
    \includegraphics[width=8cm,height=5.2cm]{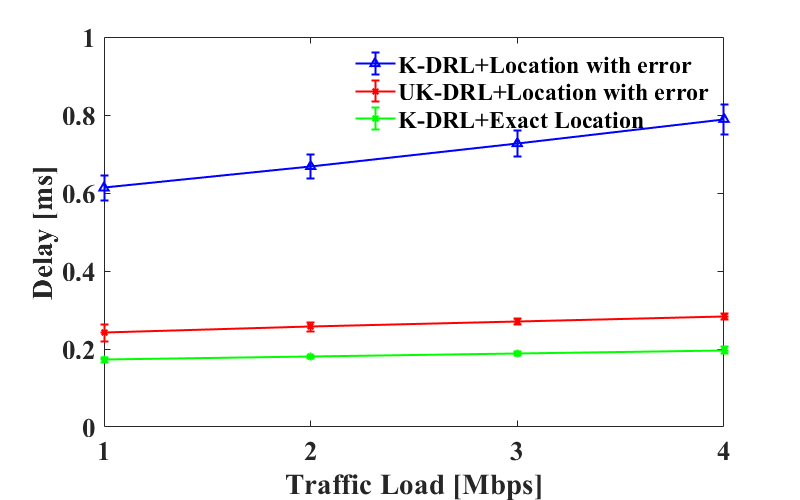}
    \caption{Delay comparison under varying traffic loads.
    \label{fig6}
}
\end{figure}

Fig. \ref{fig5} and Fig. \ref{fig6} present the sum rate and delay under different traffic loads. As shown in the plots, scenario 3 has the best performance. In scenario 3, since the gNB has the exact position of UEs and the beams are formed according to exact locations, the gNB can cover most of the UEs, leading to the highest sum rate and lowest latency. Moreover, when location error exists, UK-means proves its superiority over K-means in both sum rate and delay. In scenarios 1 and 2, gNB only has the knowledge of location with error, and the distorted positions are used to identify the direction of beams, which will degrade the system performance. However, our proposed UK-DRL can mitigate the effect of localization uncertainty by using the PDF, which will provide a more accurate clustering result, and consequently it achieves a better network performance. For instance, when the traffic load is 4 Mbps, UK-DRL has a 150$\%$ improvement in sum rate and 61.5$\%$ improvement in latency.

However, as the traffic load increases, it will stress all three algorithms. Another interesting point is that for each scenario, the delay has not changed much as the traffic load grows. 
We assume UEs to move every 10 TTIs. For the UE that is not covered, after at least 10 TTIs, it will move to a new place and is very likely to be covered.


\section{Conclusion}
\label{s6}
Machine learning has become a favourable technique to enable 5G and beyond networks. In this paper, we proposed a UK-means-based clustering and deep reinforcement learning-based resource allocation algorithm for the joint beamforming and radio resource allocation of 5G mmWave networks with localization uncertainty. In particular, we deploy the UK-means algorithm for clustering with localization uncertainty, and long short-term memory-based deep reinforcement learning for the radio resource allocation for each beam. The UK-means-based clustering and deep reinforcement-based resource allocation algorithm is compared with K-means-based clustering and deep reinforcement learning-based resource allocation algorithm, and the simulations show that UK-means-based clustering and deep reinforcement learning-based resource allocation algorithm results in higher data rate and lower delay for network users. In the future, we aim to involve more clustering algorithms to handle the localization uncertainty.

\section*{Acknowledgment}
This work is supported by Ontario  Centers  of  Excellence(OCE) 5G ENCQOR program and Ciena, the Natural Sciences and Engineering Research Council of Canada (NSERC), Collaborative Research and Training Experience Program (CREATE) under Grant 497981, and Canada Research Chairs Program. We would like to thank Medhat Elsayed and Hind Mukhtar for their help in the earlier versions of the implementations.

\vspace{12pt}


\begin{thebibliography}{00}
\bibitem{b1} V. W. Wong, R. Schober, D. W. K. Ng, and L.-C. Wang, Key Technologies for 5G Wireless Systems. Cambridge University Press, 2017.
\bibitem{b2} D. Zhang, Z. Zhou, C. Xu, Y. Zhang, J. Rodriguez and T. Sato, "Capacity Analysis of NOMA With mmWave Massive MIMO Systems," in IEEE Journal on Selected Areas in Communications, vol. 35, no. 7, pp. 1606-1618, July 2017.
\bibitem{B2} J. Cui, Z. Ding and P. Fan, "The Application of Machine Learning in mmWave-NOMA Systems," in IEEE Vehicular Technology Conference (VTC Spring), pp. 1-6, June 2018.
\bibitem{b4} C. D. McGillem and T. S. Rappaport, “Infra-red location system for navigation of autonomous vehicles,” in IEEE International Conference on Robotics and Automation, vol. 2, pp. 1236–1238, April 1988.
\bibitem{B4_3} M. Elsayed and M. Erol-Kantarci, "AI-Enabled Future Wireless Networks: Challenges, Opportunities, and Open Issues," IEEE Vehicular Technology Magazine, vol. 14, no. 3, pp. 70-77, September 2019.
\bibitem{B4_4} H. Zhou, M. Elsayed, and M. Erol-Kantarci, “RAN Resource Slicing in 5G Using Multi-Agent Correlated Q-Learning,” in Proceedings of 2021 IEEE conference on PIMRC, pp.1-6, Sep. 2021.
\bibitem{B_zh} H. Zhang, H. Zhou, M. Erol-Kantarci, “Team Learning-Based Resource Allocation for Open Radio Access Network (O-RAN),” arXiv:2201.07385, pp.1-6, Jan. 2022.
\bibitem{b5} M. Chau, R. Cheng, B. Kao, and J. Ng, “Uncertain data mining: An example in clustering location data,” in Pacific-Asia conference on knowledge discovery and data mining, vol.2, pp. 199-204, April 2006.
\bibitem{b6} M. Elsayed and M. Erol-Kantarci, "Radio Resource and Beam Management in 5G mmWave Using Clustering and Deep Reinforcement Learning," in IEEE Global Communications Conference (GLOBECOM), pp. 1-6, December 2020.
\bibitem{b7} D. Marasinghe, N. Jayaweera, N. Rajatheva, and M. Latva-Aho, “Hierarchical User Clustering for mmWave-NOMA Systems,” in 6G Wireless Summit (6G SUMMIT), pp. 1–5, February 2020.
\bibitem{b8} I. Orikumhi, H. -K. Jwa, J. -H. Na and S. Kim, "Location-Aided User Clustering and Power Allocation for NOMA in 5G mmWave Networks," in International Conference on Information and Communication Technology Convergence (ICTC), pp. 264-268, October 2020.
\bibitem{b9} J. Cui, Z. Ding, P. Fan, and N. Al-Dhahir, “Unsupervised Machine Learning-Based User Clustering in Millimeter-Wave-NOMA Systems,” IEEE Transactions on Wireless Communications, vol. 17, pp. 7425–7440, November 2018.
\bibitem{b12} A. Alkhateeb, G. Leus and R. W. Heath, "Limited Feedback Hybrid Precoding for Multi-User Millimeter Wave Systems," in IEEE Transactions on Wireless Communications, vol. 14, no. 11, pp. 6481-6494, November 2015.
\bibitem{B12_1} A. Klautau, P. Batista, N. González-Prelcic, Y. Wang and R. W. Heath, "5G MIMO Data for Machine Learning: Application to Beam-Selection Using Deep Learning," in Information Theory and Applications Workshop (ITA), pp. 1-9, February 2018.
\bibitem{B12_2} H. Mukhtar, Melike Erol-Kantarci, “Machine Learning-Enabled Localization in 5G Using LIDAR and RSS Data,” in Proc. of IEEE ISCC, September 2021.
\bibitem{b13} N. I. B. Hamid, N. Salele, M. T. Harouna, and R. Muhammad, “Analysis of LTE Radio Parameters in Different Environments and Transmission Modes,” International Conference on Electrical Information and Communication Technology (EICT), pp. 1–6, March 2014.
\end{thebibliography}
\end{document}